# Metallic spintronic nanofilm as a hydrogen sensor

Crosby S. Chang,[1] Mikhail Kostylev,[1*] and Eugene Ivanov[1]

[1]*School of Physics M013, The University of Western Australia, 6009 Crawley, WA, Australia*

Abstract: We investigate the response of palladium-cobalt bi-layer thin films to hydrogen charging at atmospheric pressure for spintronic applications. We find that hydrogen absorption by the palladium layer results in the narrowing and shifting of the ferromagnetic resonance line for the material. We explain the observed phenomena as originating from reduction in spin pumping effect and from variation in the magnetic anisotropy of the cobalt film through an interface effect. The shift of the resonance frequency or field is the easiest to detect. We utilize it to demonstrate functionality of the bi-layer films as a hydrogen sensor.

Recently, the demand for highly stable and sensitive hydrogen gas sensors has increased due to growing importance of fuel cell applications. Various types of hydrogen gas sensors have been proposed [1]. All the proposed sensors have a number of drawbacks. In particular, sensors either based on variation of resistance of semiconductors [2], or on catalytic oxidation of noble metals

---

[*] Corresponding author. Email address: mikhail.kostylev@uwa.edu.au

[3] have poor selectivity for hydrogen gas and require heating the sensor head. Optical [4] and micro-mechanical sensors [5] require sophisticated optical detection systems. An interesting concept has been suggested quite recently, utilizing a palladium/organic ferroelectric multilayer film to sense the presence of hydrogen [6]. The main drawback of this and of many other concepts concept is the fire hazard associated with the need to generate or apply electrical potentials directly in the hydrogen environment. The novel concept of plasmonic sensors [7-10] is free from this drawback, but also requires a sophisticated optical detection method. Furthermore, it requires that the vessel containing the monitored volume has a transparent wall [7, 9-10].

Many of the suggested hydrogen sensors make use of the well-known property of palladium to absorb and chemically bind hydrogen [1,5-10]. It is less known that palladium also has unique spintronic properties. Magnetic multilayered nano-films and superlattices which include non-magnetic palladium layers are of very large interest for the magnetic community because of their importance for new concepts of high-density magnetic random access memory utilizing nanoscale magnetic tunnel junctions [11]. This interest stems from the strong perpendicular magnetic anisotropy (PMA) demonstrated for Co/Pd [11-12] and CoFe/Pd [13] multilayers on which these concepts rely on. Palladium [14] and similar hydrogen-sensitive niobium non-magnetic metallic spacers [15] have also been used in magnetic spin valve nanostructures which are important to numerous spintronic applications, including applications in reading heads for modern hard drives. Charging spacers with hydrogen at high pressure led to variation of exchange coupling between magnetic layers in these devices [15]. Furthermore, palladium layers overlaying magnetic layers demonstrate a very large inverse spin Hall effect [16] which is important for microwave magnonic applications [17] Palladium/magnetic metal bi-layers also show significant spin-pumping effect [16, 18-20] , Interestingly, there have been investigations of hydrogen charging of multi-layers containing palladium and magnetic metals for applications in membranes for gas separation [21] and in optical switching [22]. Also, there were a small number of studies of variation in PMA upon absorption of Pd/Co superlattices with hydrogen. The study [23] demonstrated an initial increase in PMA with hydrogen absorption and then a drop for larger partial hydrogen gas pressures. Unfortunately, this study was carried out for partial pressures well above 1 atmosphere (1.3-5.2 atm). The works [28,29] reported a decrease in PMA on hydrogen absorption. Methods which are very impractical for sensor applications - vibrating sample magnetometry and polarized neutron reflectivity - were utilized in these works. There was also a magneto-optical study of a Pd/Co film in hydrogen-containing atmosphere [30]. The authors registered variation of optical properties of Pd with hydrogen absorption, but found no evidence of the variation of the magnetic properties of the ferromagnetic layer upon hydrogen absorption.

In this letter, we present a different concept of a hydrogen sensor which may sense the presence of hydrogen remotely, as optical sensors do, but does not require optical readout. Furthermore, it can also sense hydrogen through an optically non-transparent (but still electrically insulating) wall. The sensor we suggest is based on modification of magnetic properties upon hydrogenation of a palladium layer in non-magnetic metal/magnetic metal (NM/MM) bi-layer nanomaterials which are widely used in spintronic applications. The palladium layer plays the role of the NM layer in these nanofilms. In contrast to [15] where the functionality of their device requires increased hydrogen pressure, the results we demonstrate here were obtained at atmospheric pressure. We employ the ferromagnetic resonance (FMR) method to probe this modification.

Ferromagnetic resonance represents an eigen-excitation in magnetic materials which exists in the microwave frequency range and is seen as resonant absorption of microwave power by a magnetic sample when the frequency of a microwave source is equal to the resonance frequency of the material. The FMR frequency is given by the material parameters and also depends on the external magnetic field applied to the sample. We will use the FMR frequency dependence on external applied magnetic field in our demonstration of the functionality of this device. An extremely low amplitude microwave field can penetrate an electrically insulating wall and probe the sensor through the wall. More importantly, because the sensor is made from metallic conductors, the sensor itself serves as a perfect shield for microwaves [27-28], not allowing the microwave power to penetrate into the vessel that potentially contains the hydrogen gas. Thus, there will be no electrical fields capable ignite the gas present inside the vessel.

To demonstrate our concept, we fabricated a number of bi-layer non-magnetic metal/magnetic metal films in which palladium plays the role of the non-magnetic layer. The films were obtained with magnetron sputtering. Two materials were used for the magnetic layers: cobalt and Permalloy (NiFe); the latter is a metallic alloy with composition 80% nickel and 20% iron. First, a 5 nm thick tantalum seed layer is sputtered onto a 0.9 mm thick silicon substrate. Then, the magnetic layer (NiFe or Co) were sputtered on top of the seed layer. Finally, the palladium layer was sputtered on top of the magnetic layer. Several films with different thicknesses of palladium and magnetic layers have been produced. They are listed in Table I.

| NiFe/Pd bi-layers | NiFe (5)/Pd(10) |
|---|---|
| | |

| Co/Pd bi-layers | NiFe (30)/Pd(10) |
|---|---|
| | Co(5)/Pd(10) |
| | Co(40)/Pd(20) |

Table I. Bilayers fabricated, where the numbers in brackets indicate the film thickness (nm).

A custom 4 cm x 4 cm x 4 cm cubic air-tight cell (FIG. 1) was made to enable controlled continuous flow of gas at atmospheric pressure through the chamber while performing FMR experiments. The cell contains a coplanar waveguide (CPW) on which the samples sit. Coaxial cables feed microwave power from a microwave generator into the CPW from one end, and carry the transmitted power out from another end into a microwave receiver. The cell is fixed between the poles of an electromagnet such that the magnetic field is applied along the CPW. This orientation maximizes the FMR response.

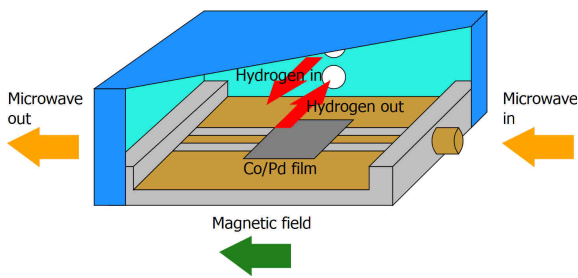

FIG. 1. (Color) Cross-section of the gas cell showing the coplanar waveguide (on which the sample sits on), microwave feed ports, and gas flow inlets.

We used the standard field-modulated (FM) ferromagnetic resonance method to take the measurements. The typical FM-FMR experiment run is as follows [29]. The sample is placed on top of the CPW line which forms the bottom of the fixture. The cell is sealed air-tight, and gas introduced into the chamber (nitrogen or hydrogen). Microwave power of a specific frequency (typically in the range of 4 – 18 GHz) is applied to the CPW. The output end of the CPW is connected to a microwave receiver. The rectified signal from the receiver output is fed to a lock-in amplifier. An extra ac field of ~ 1 Oe rms at 220 Hz is applied to the sample using the modulating coil; this is the signal referenced by the lock-in amplifier. The dc magnetic field is swept for each particular microwave frequency. The FMR absorption then represents a dip in the microwave power transmitted through the CPW line as a function of the applied field. Due to field modulation, the resultant signal detected by the lock-in amplifier is the first derivative of the resonance curve as a function of sweeping field at the set microwave frequency (differential absorption). Experiment was carried out in both nitrogen and hydrogen atmospheres at atmospheric pressure, and the FMR traces compared.

For the thicker films – NiFe(30)/Pd(10) and Co(40)/Pd(20) – we did not observe any appreciable differences in the FMR spectra upon hydrogen charging of the palladium layer. For the thinner samples, we do observe the effect of hydrogen charging on the FMR response (FIG. 2). The strong dependence of the effect on the thickness of the magnetic layer suggests that the effect is of interfacial nature. Hence, for the remainder of this paper, we now turn our attention to the thinner films; NiFe(5)/Pd(10) and Co(5)/Pd(10).

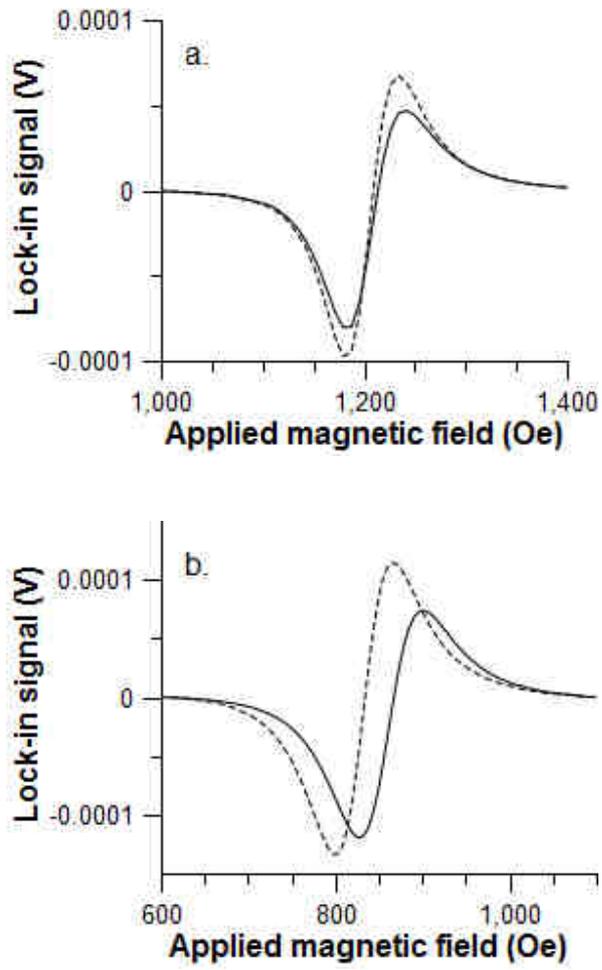

FIG. 2. Raw FMR spectra at 10 GHz in nitrogen atmosphere (solid line) and hydrogen atmosphere (dashed line). (a) NiFe(5)/Pd(10)  (b)  Co(5)/Pd(10) .

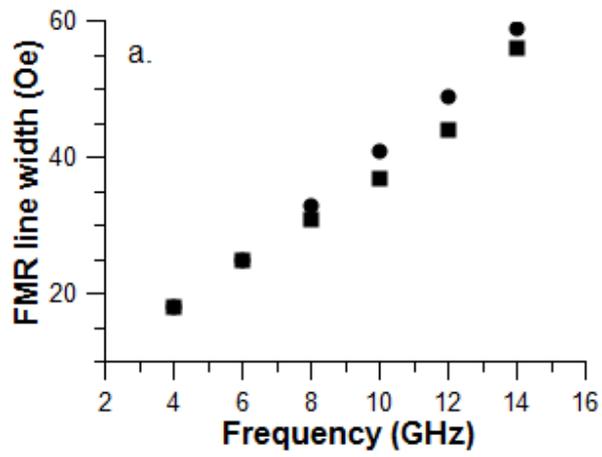

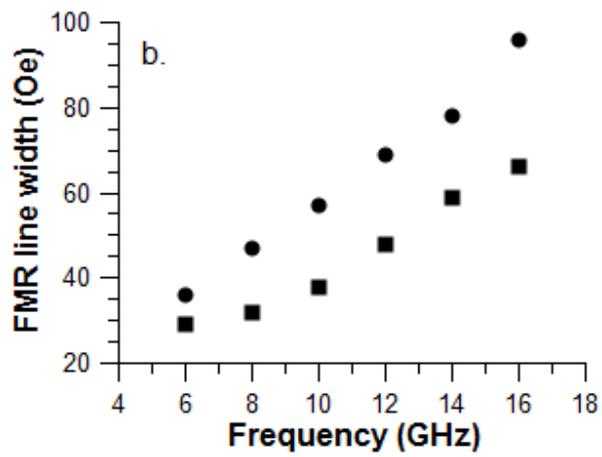

FIG. 3. FMR linewidth at various frequencies in nitrogen atmosphere (circles) and hydrogen atmosphere (squares). (a) NiFe(5)/Pd(10) (b) Co(5)/Pd(10).

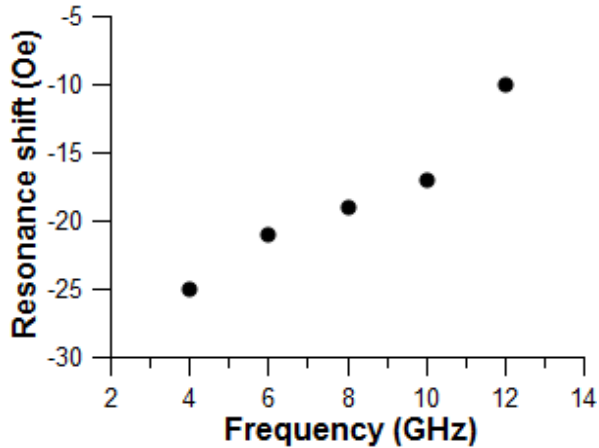

FIG. 4. FMR frequency shift upon hydrogenation of Co(5)/Pd(10) at various frequencies.

By fitting the raw experimental data (FIG. 2) with the first derivative of the Cauchy-Lorentz function, one can extract the three parameters which define resonance: position, width and amplitude. For the NiFe(5)/Pd(10) film (FIG. 3a), upon hydrogenation of the palladium layer, the resonance line width decreased while the amplitude increased. For the Co(5)/Pd(10) film, the resonance line width narrowing and amplitude increase is noticeably larger upon hydrogenation (FIG. 3b). In addition, one immediately notices that the resonance line shifted down field upon hydrogenation (FIG. 2b). It is worth noting these shifts are larger than the experimental uncertainty in the magnetic field (~ 5 Oe). FIG. 4 shows the resonance frequency shift at various frequencies.

Thus, the largest effect caused by hydrogenation is the shift of the resonance line for the Co(5)/Pd(10) sample. As mentioned above, there is strong evidence that the effect is of interfacial nature. A typical magnetic interfacial effect is perpendicular anisotropy induced in a number of FM/NM multilayers [11-13]. Co/Pd is a typical material with perpendicular magnetization [11-12]; where uniaxial anisotropy at the interface of the two layers is strong enough to force the magnetization vector out of plane, provided the film is sufficiently thin. However, our layers are too thick to force the magnetization vector out of plane; the magnetization vector in the Co layer remains in-plane. Therefore the shift in the resonance field cannot be attributed, for instance, to switching of equilibrium magnetization from out-of-plane to in-plane upon hydrogen charging. A more plausible explanation is that the interface uniaxial anisotropy changes when hydrogen enters palladium layer. It is known that the origin of the perpendicular anisotropy in Co/Pt-group NM multilatyers is an interface strain [30]. It also known that palladium expands on hydrogen charging [6] which should result in variation of the strain and thus variation of magnetic anisotropy in cobalt. The change in the magnitude of the anisotropy field shifts the resonance frequency. This conclusion is consistent with a much smaller effect we see for the NiFe/Pd bi-layer, since NiFe is a material with negligible anisotropy and magnetostriction.

As for the decrease in the resonance line width upon hydrogen charging, there are two possible contributions to this effect. The first one is the spintronic effect of spin pumping which is also of interfacial nature. It is known palladium is one of the materials in which this effect is strong [20]. The absorption of hydrogen reduces the conductivity of the palladium layer, which results in reduction in spin-pumping contribution to the resonance line width due to reduced spin-mixing conductance at the interface [31].

The second effect is the trivial effect of reduction of the contribution of the eddy current losses to the resonance line upon reduction in the palladium layer conductivity. To estimate this contribution, we performed simulations of the microwave response of the CPW loaded by a NM/MM sample for different values of conductivity for the NM layer [27]. Note that this simulation does not take into account the spin pumping effect. We found that reduction in the conductivity of the NM layer from the one typical for bulk palladium to zero has a negligible effect on the resonance line width. Hence, we conclude that spin pumping into the non-magnetic palladium layer is the dominating contribution to resonance line broadening.

Next, we exploit the dependence of the resonance frequency shift in Co(5)/Pd(10) upon hydrogen charging for demonstration of functionality of our device as a hydrogen sensor.. To this end we cycle nitrogen and hydrogen atmospheres through the chamber. First, we set the frequency and field to ones within the FMR line. We do this in nitrogen atmosphere. In this particular trial, we set the frequency to 10 GHz and field to 867 Oe; this is one particular resonant condition for the Co(5)/Pd(10) film. The cell atmosphere was then alternated between nitrogen and hydrogen cyclically three times. The output voltage of the lock-in amplifier is measured as a function of time with a digital oscilloscope during these cycles. FIG. 5 shows the results from this cyclic run.

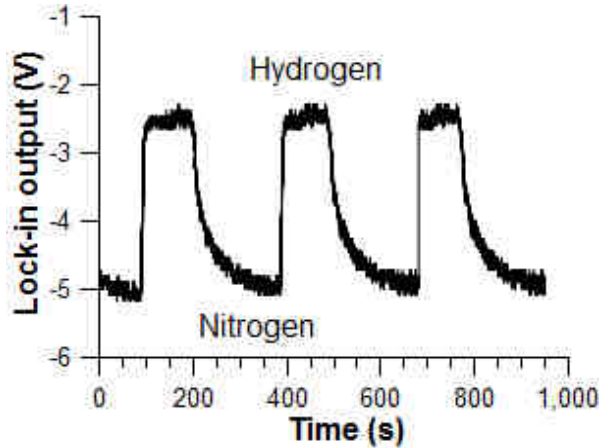

FIG. 5. Change in the lock-in output voltage under the cycling of nitrogen and hydrogen gas through the Co(5)/Pd(10) film under resonance conditions.

Since the frequency and field were set to resonance conditions under nitrogen atmosphere, the change in the signal baseline upon the introduction of hydrogen is due to the cobalt layer going out of resonance conditions. Cycling of nitrogen and hydrogen thus results in the system going in and out of resonance, respectively. This is easily detected by the lock-in amplifier.

We note three key features from this cyclic run which are ideal in any sensor. First, the baseline change due to the sensing of hydrogen is well above noise level. Second, each cycle is reproducible; that is, the system reliably returns to the same state in each cycle. Third, the sensor response and reset times are reasonably fast. The hydrogen sensing and reset time constants were found to be 5s and 30s respectively. These values are similar to the response times of a typical electrical resistance based palladium thin film hydrogen sensor [32-33]. This further confirms that the signal is due to absorption and desorption of hydrogen at the palladium, rather than gas flow artefacts.

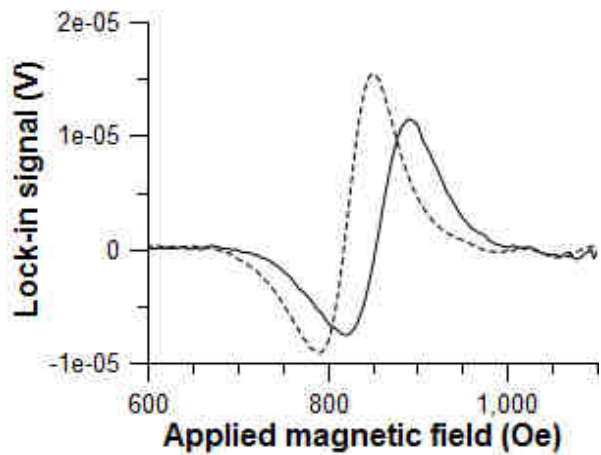

FIG. 6. FMR spectra of flipped Co(5)/Pd(10) with the metallic film facing away from the CPW at 10 GHz in nitrogen atmosphere (solid line) and hydrogen atmosphere (dashed line).

The last stage of the experiment is the demonstration of the possibility of remote sensing of hydrogen. In all previous experiments the sample was placed on the CPW such that the metal film faces the transducer. In this experiment we flip the sample such that the bi-layer film now faces away from the CPW (FIG. 6). Thus, the film is now separated from the coplanar line by the mm-thick insulating silicon substrate. This arrangement mimics a vessel wall between CPW attached to the external surface of the wall and the film on the internal surface of the wall.

As one sees from (FIG. 6) the resonance is easily detected in this configuration through an electrically insulating wall even though the signal has dropped by 20 dB (see FIG. 2b and compare with FIG. 6). Importantly, as follows from references [27-28], the microwave field in this configuration is concentrated in the insulator and the metal film and penetration into the vessel is negligible due to the perfect microwave shielding effect exhibited by metallic films of sub-skin-depth thicknesses.

In conclusion, we demonstrated that a metallic Co/Pd bi-layer nanofilm largely used for spintronic applications also has the functionality of a hydrogen sensor. Hydrogen charging of the palladium layer results in modification of two interfacial effects: a.) it varies the magneto-crystalline anisotropy in cobalt through variation in interfacial stress and b.) reduces microwave magnetic losses in cobalt through reduction in spin pumping effect. Both effects can be detected in a conventional microwave ferromagnetic resonance absorption experiment. The resonance frequency shift is the easiest to detect. The shift of the resonance frequency upon charging the palladium layer with hydrogen results in a variation of the amplitude of the microwave signal from the output of the sensor. This variation is repeatable as demonstrated by cycling between nitrogen and hydrogen atmospheres. We found that the response and reset times of this sensor are typical for such thin film palladium hydrogen sensors. We also demonstrated that the microwave method allows sensing of the hydrogen hazard remotely through a non-transparent 1 mm-thick insulating wall.

There is one more aspect which also needs commenting. This is the application of the magnetic field in our experiment. The field is usually needed to magnetically saturate the sample. The saturated or single-domain state is an important condition for observation of FMR. We speculate that in real sensors, instead of continuous films, one may use films in which the magnetic layer is not continuous, but is nanopatterned, similar to optical sensors [7]. For instance, one can form an array of magnetic nanostripes instead of a continuous film and fill in the spaces between the stripes with some nonmagnetic metal [34] which is not sensitive to the presence of hydrogen. The nanostripes are naturally in single-domain state and FMR in them is observable without application of a magnetic field [35]. Filling in the gaps between the stripes with a non-magnetic metal will create a continuous film below a hydrogen sensitive palladium layer. This will preserve the perfect microwave shielding effect.

ACKNOWLEDGMENT

Financial support from by the Australian Research Council is acknowledged.